\documentclass[pra,twocolumn,aps]{revtex4-1}

\usepackage{graphicx}
\usepackage[colorlinks=true,linkcolor=blue,citecolor=red,urlcolor=blue]{hyperref}
\usepackage{amsmath, amssymb}
\usepackage{bm}
\usepackage{natbib}
\usepackage{color}
\usepackage{soul,xcolor}

\def\mbf{\mathbf}
\newcommand{\be}{\begin{equation}}
\newcommand{\ee}{\end{equation}}

\begin{document}
\title{Rotating black hole geometries in a two-dimensional photon superfluid.} 
\author{David Vocke$^{1}$, Calum Maitland$^{1}$, Angus Prain$^{1}$, Fabio Biancalana$^{1}$, Francesco Marino$^{2,3}$, Ewan M. Wright$^{1,4}$, Daniele Faccio$^{1}$}
\email{d.faccio@hw.ac.uk, d.vocke@hw.ac.uk}
\affiliation{$^1$Institute of Photonics and Quantum Sciences, Heriot-Watt University, Edinburgh EH14 4AS, UK} 
\affiliation{$^2$CNR-Istituto Nazionale di Ottica, L.go E. Fermi 6, I-50125 Firenze, Italy}
\affiliation{$^3$INFN, Sez. di Firenze, Via Sansone 1, I-50019 Sesto Fiorentino (FI), Italy}
\affiliation{$^4$College of Optical Sciences, University of Arizona, Tucson, AZ, USA}

\begin{abstract}
Analogue gravity studies the physics of curved spacetime in laboratory experiments, where the propagation of elementary excitations in inhomogeneous flows is mapped to those of scalar fields in a curved spacetime metric. While most analogue gravity experiments are performed in 1+1 dimensions (one spatial plus time) and thus can only mimic only 1+1D spacetime, we present a 2+1D photon (room temperature) superfluid where the geometry of a rotating acoustic black hole can be realized in 2+1D dimensions.
By measuring the local flow velocity and speed of waves in the superfluid, we identify a 2D region surrounded by an ergo sphere and a spatially separated event horizon. This provides the first direct experimental evidence of an ergosphere and horizon in any system, and the possibility in the future to study the analogue of Penrose superradiance from rotating black holes with quantised angular momentum and modified dispersion relations.
\end{abstract}

\maketitle
{\textit{Introduction.}}---
The field of analogue gravity  {has} demonstrated that the physics of distorted spacetime can be studied in laboratory environments that exploits the formal analogy between waves in inhomogeneous fluid flows and scalar fields in curved spacetime \cite{Unruh1981,Visser1998, Barcelo2005,Faccio2013}. In this context, a spatially varying flow maps onto an effective spacetime metric in which perturbative excitations, i.e. density or surface waves propagate. A horizon is formed  {at a surface} where the flow speed  {across that surface} exceeds  {the} wave propagation speed and hence waves are blocked or trapped beyond that boundary.

Analogue horizons have been realised in various systems, where quantum fluids such as Bose-Einstein Condensates (BECs) and classical systems such as surface waves in water or pulses of light in an optical medium play the most prominent roles \cite{Leonhardt2008,Belgiorno2010,Weinfurtner2011,Lahav2010,Steinhauer2015,Steinhauer2016,Jacobson1998}.  These experimental studies involved one dimensional flow geometries and have had considerable success. The challenging realisation of higher dimensional analogue horizons is  {only recently being undertaken} \cite{Torres2017} and would enable  {the} study  {of the} effects of rotating spacetimes  {for example} by setting the background flow into rotation similar to a vortex in a draining bathtub. 

A rotating spacetime may form an ergosphere enclosing the internal region within which it is impossible for an observer to remain stationary relative to a distant observer. Recent experiments using draining water tanks showed that waves entering a hydrodynamic vortex may be scattered and amplified \cite{Torres2017}, effectively extracting rotational energy.  {This effect is related to a particle scattering effect first predicted by Penrose in 1969 \cite{Penrose1969, Penrose1971} referred to as the Penrose process and was later discussed in the context of electromagnetic waves incident on rotating conducting cylinders \cite{Zeldovich1972} and in the context of rotating spacetimes \cite{staro1,staro2} (see also \cite{ Bekenstein1998, Cardoso2017,Endlich2017,Visser2005,Richartz2013,Basak2003,Marino2008, Marino2009,berti, lepe,ednilton}). 
In a quantum fluid however, where vortices and thus the associated background rotating spacetime are inherently quantized, such an effect has not been experimentally studied yet. Related to this, the experimental realization of a 2+1D vortex flow possessing both a horizon and ergosphere has never been achieved so far.

Recently, a new approach to quantum  fluids has emerged in the form of quantum  fluids of light, where effective photon-photon interactions of a monochromatic laser beam propagating in a nonlinear medium lead to a collective behaviour of the many photon system, leading to a superfluid behaviour \cite{Frisch,Chiao1999,Vocke2015,Vocke2015a}.
The fluid density,  {which} defines the speed of  {linear excitations} (referred to here as ``sound'' waves) is determined by the laser intensity while the overall flow is controlled via (the gradient of) the spatial phase profile, making these systems a promising platform to tailor-make  {2+1D} analogue spacetimes \cite{Marino2009, Marino2009}.
 
In this work, we use a room-temperature photon superfluid to create a  {2+1D} {rotating} spacetime with an inward,  {draining} radial flow. Measurements of the fluid density (i.e. wave speed) and phase gradients (i.e. flow speeds) along the radial and angular directions allow us to precisely identify the horizon and ergosphere  {location} in the superfluid.   We also show that by inverting the superfluid phase topology, a rotating white hole, the time reversal of a black hole, is created. 
By mapping out the spacetime metric through measuring the flows we calculate derived geometric quantities such as the surface gravity of the effective horizon, the angular velocity of the horizon and the size of the ergo-region.  


{\textit{The model.}}---
Our photon fluid involves a monochromatic laser beam that propagates through  {a} bulk medium with a thermo-optic nonlinearity \cite{Ghofraniha2007, Conti2009}. In the paraxial approximation, the slowly varying envelope of the electric field $E(\mbf{r},z)$ with $\mbf{r}=(x,y)$ is governed by the Nonlinear Schr\"odinger Equation (NLSE):
\begin{equation}
\partial_zE=\frac{i}{2k}\nabla_{\perp}^2E+i\frac{k}{n_0}\Delta nE,
\label{eq:NLSE}
\end{equation}
where $z$ is the propagation direction and $k=2\pi n_0/\lambda$ is the  {longitudinal} wave number. The nonlinear change of refractive index  due to the thermo-optic response is described by 
\begin{equation}
\Delta n=\gamma\int\int dr'dz' R(r-r',z-z')|E(r',z')|^2,
\end{equation}
where $\gamma$ is the nonlinearity coefficient and $R(r,z)$ is the nonlocal response function that will typically have an exponential-like decay in the transverse, radial direction with spatial extent $\sigma$ that accounts for heat diffusion resulting from the absorbed laser power. If heat diffusion is sufficiently limited i.e. $\sigma<\Lambda$ where $\Lambda$ is the sound wavelength, then the nonlocal response function can be approximated as local, $R(r-r',z-z')=\delta(r-r',z-z')$ and thus $\Delta n=\gamma |E(r,z)|^2$. We will later see that this is a valid assumption for the length scales considered in the experiments.

For a defocusing nonlinearity, i.e. $\Delta n<0$, the NLSE can be recast in a set of hydrodynamical equations by means of the Madelung transform $E(r,t)=\sqrt{\rho(r,t)}e^{i\phi(r,t)}$   \cite{Chiao1999,Carusotto2014}, as also found for dilute Bose-Einstein condensates (BECs) \cite{Dalfovo1999}
\begin{align}
\partial_t\rho+\nabla(\rho v)&=0 \label{eq:conti} \\ 
\partial_t\psi+\frac{1}{2}v^2+\frac{c^2\gamma}{n_0^3}\rho-\frac{c^2}{2k^2n_0^2}\frac{\nabla^2\sqrt{\rho}}{\sqrt{\rho}}&=0 \label{eq:euler}
\end{align}
\begin{figure}[t!]
\begin{center}
\includegraphics[width=8cm]{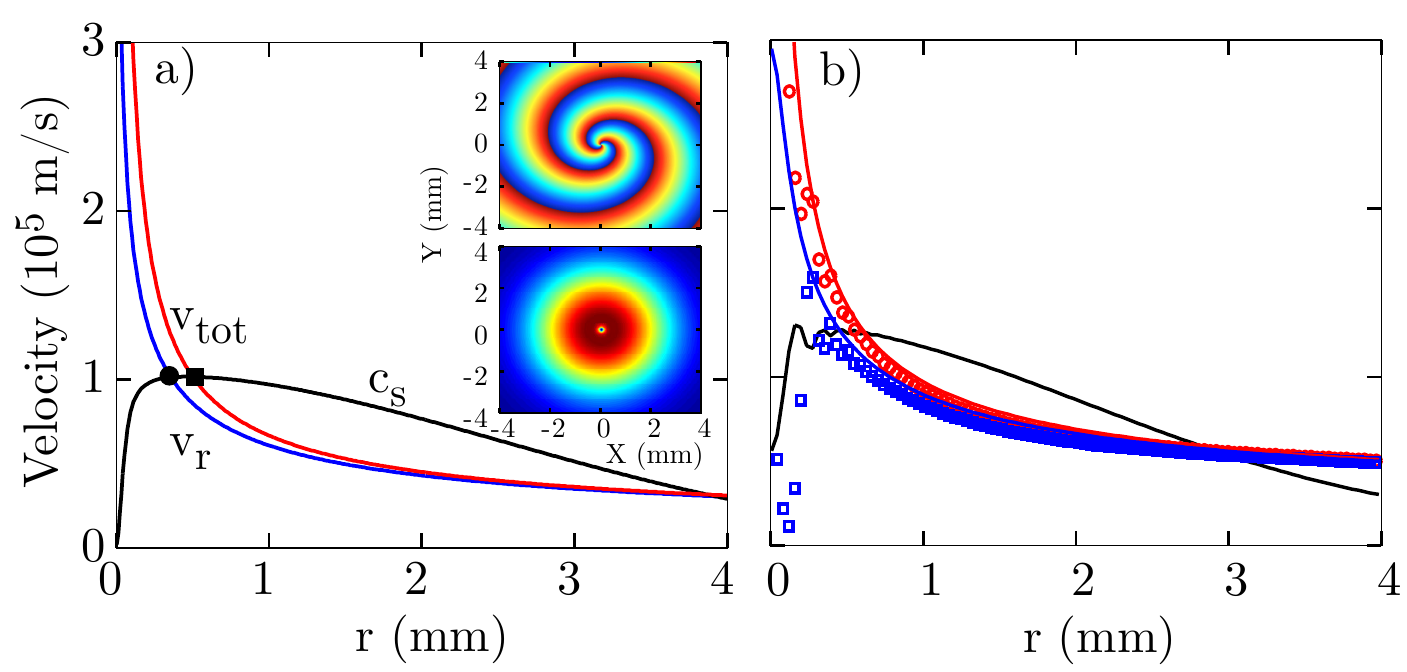}
\caption{a) Flow and sound wave velocities calculated for a photon fluid  with gaussian intensity envelope (width $w_{1/e^2}$=5 mm, P=140 mW, $|\gamma|=4.4\times10^{-7}$ $\mathrm{cm^2/W}$) and field amplitude $E_0=\sqrt{\rho_0(r)}\exp{(im\theta-2i\pi\sqrt{r/r_0})}$ with $r_0=0.5$ mm and m=2. The solid circle(square) indicates the location of the horizon(ergosphere). Inset top: 2D Phase profile with values from 0 (blue) to 2$\pi$ (red). Inset bottom: Nearfield intensity profile (arb. uts). b) Flow and sound wave velocities after 13 cm propagation in a nonlocal nonlinear medium with the same parameters obtained from numerical integration of the NLSE [Eq.~(\ref{eq:NLSE})]. }
\label{fig:BHcalculated}
\end{center}
\end{figure}
where the propagation axis is mapped onto  {a} time coordinate by way of $t=zn_0/c$, the transverse phase gradient determines the fluid flow $\mbf{v}(r,t)=(c/kn_0)\nabla_{\perp}\phi(r,t)=\nabla_\perp\psi(r,t)$ and the laser intensity determines the speed of sound $c_s(r,t)=\sqrt{c^2|\gamma|\rho(r,t)/n_0^3}$. Therefore in the presence of repulsive nonlinear interactions, the transverse beam profile follows the mean-field dynamics of a Bose-Einstein condensate with the optical field $E$ playing the role of the complex order parameter. 

The equation of motion for sound waves can be obtained by linearising Eq. (\ref{eq:conti}) and (\ref{eq:euler}) around a stationary background solution, where elementary excitations are understood as first order fluctuations of the optical field amplitude and phase, i.e. $\rho=\rho_0+\rho_1$ and $\psi=\psi_0+\psi_1$ with $\rho_1\ll\rho_0$ and $\psi_1\ll\psi_0$. The last term in Eq. (\ref{eq:euler}) is the quantum pressure that appears in optics due to diffraction and can be neglected in the hydrodynamic limit that is valid for length scales much larger than the healing length, $\zeta=\lambda/\sqrt{4n_0|\gamma|\rho}$. Typically, in our system this implies sound waves with wavelengths $\geq 300$ $\mu$m \cite{Vocke2015}. 
In this case, it can be shown that the equation of motion for the phase perturbations $\psi_1$ can be reformulated into  {a} Klein-Gordon equation for a massless scalar field on a curved spacetime \cite{Marino2008}
\begin{equation}
\Box\psi_1=\frac{1}{\sqrt{-\mbf{g}}}\partial_\mu\left(\sqrt{-\mbf{g}}\,g^{\mu\nu}\partial_\nu\psi_1\right)=0
\label{eq:KleinGordon}
\end{equation}
where the spacetime metric $g_{\mu\nu}$ is given by
\begin{equation}
g_{\mu\nu}=\bigg(\frac{\rho_0}{c_s}\bigg)^2\begin{pmatrix}
						 -(c_s^2-v^2) & -v_r & - {r}v_\theta \\
						-v_r & 1& 0 \\
						- {r}v_\theta & 0 &  {r^2} \\
			\end{pmatrix}
\label{eq:photonfluidmetric}					
\end{equation}
and $\mbf{g}=\mathrm{det}(g_{\mu\nu})$.
 Here, $v_r$ and $v_\theta$ are the radial and azimuthal velocity components, from which the total speed is $\mbf{v}^2=v_r^2+v_\theta^2$. This metric is well known for classical fluids and BECs and depending on the flow geometry, can have  a horizon and a spatially separated ergosphere. Note that $v$ and $c_s$ are functions of the transverse coordinates, hence by tailoring the spatial phase and intensity profile it is possible to generate  {a wide class of} (2+1) dimensional spacetime metrics.
 
Following the initial proposal by Marino \cite{Marino2008}, a rotating spacetime may be created by using background vortex beams with orbital angular momentum (OAM) $E_0=\sqrt{\rho_0(r)}e^{im\theta}$, with topological charge integer $m$. Such beams are characterised by a phase singularity surrounded by a helical wave front: the phase gradient and therefore the azimuthal fluid flow $v_{\theta}(r)=cm/(kn_0r)$ is quantized and proportional  {as a function of $r$} to the topological charge of the beam. An ergosphere can then be created by controlling the beam intensity such that the speed of sound passes from faster to slower than the  {total} flow. Any sonic observer within this region is forced to co-rotate with the flow due to the associated ``superluminal" dragging of inertial frames. In order to create a trapped surface and thus an (apparent) horizon, an additional radial phase dependence must be imposed. In any region where the flow is inward-pointing and $\vert v_r \vert > c_s$, a sound wave will be swept inward by the fluid flow and be trapped inside the horizon, that is formed where $\vert v_r \vert = c_s$.

Figure~\ref{fig:BHcalculated}(a) shows the calculated absolute flow amplitudes and corresponding sound speed for a beam with amplitude  $E_0=\sqrt{\rho_0(r)}\exp{(im\theta-2i\pi\sqrt{r/r_0})}$, with gaussian intensity envelope $\rho_0(r)=\rho_0\exp(-2r^2/\sigma^2)$ (which, upon propagation will form a ring-shaped beam due to the phase singularity at $r=0$). Figure~\ref{fig:BHcalculated}(b) shows the same input beam propagated over a distance of 13 cm by numerically solving the NLSE using a split-step Fourier method (details of this standard numerical method can be found in \cite{Vocke2015a}). 

From the flows and the metric \eqref{eq:photonfluidmetric} one can then calculate two derived quantities which are of central importance for sonic propagation on a curved background: the surface gravity and rotational velocity of the horizon \cite{Richartz2015}. The analogue surface gravity is a measure of the effective mass of the black hole, which fixes a scale for processes that occur at or near the horizon and is given by \cite{Barcelo2005}
\be
\kappa:=\left. \frac{1}{2}\partial_r\left(c_s^2-v_r^2\right)\right|_{\text{horizon}}.
\ee
For the parameters used in Fig.~\ref{fig:BHcalculated}, the average of the initial and final values provide $\langle\kappa\rangle\simeq 1.74\times 10^{13}$ ms${}^{-2}$ which corresponds to phonons of wavelength $\lambda= c_s^2/\kappa \simeq 1$ mm that are longer than the healing length of 300 $\mu$m and hence in the sonic regime. This parameter also occupies a central role in the Hawking process for analogue black holes \cite{Paren1,Paren2}.  Conversely, in superradiant scattering the angular velocity of the horizon is the most relevant parameter, setting the frequencies of the most efficiently superradiated modes through 
\be
\omega_\text{super}\simeq \frac{v_\theta(r_H)}{r_H}
\ee
where $r_H$ is the location of the horizon (solid circle in Fig.~\ref{fig:BHcalculated}). Again for the parameters of Fig.~\ref{fig:BHcalculated}, we find the average $\langle\omega_{\text{super}}\rangle\simeq 2.03\times 10^8$ s ${}^{-1}$, which corresponds to phonons with $\lambda=2\pi c_s/\omega_\text{super}\simeq 1$ mm well within the sonic regime . 

We can also estimate the time scale of the evolution of the surface gravity as a measure of the evolution of the background through the experiment,
\be
\tau^{-1} :=\frac{\dot{\kappa}}{\kappa}\simeq\frac{|\kappa_f-\kappa_i|}{\Delta t}\frac{1}{\langle\kappa\rangle}.
\ee
Using the data of Fig.~\ref{fig:BHcalculated} we find $\tau^{-1}\simeq 1.04\times 10^{9} $ s${}^{-1}$ corresponding to a  wavelength of $\lambda=2\pi c_s \tau\simeq 1$ mm: shorter wavelength phonons will perceive the background to evolve adiabatically. This shows that under realistic experimental conditions, it is indeed possible to create a spacetime vortex flow in a photon fluid and that this is maintained (i.e. is stationary) over the relevant sonic time scales.  \\
Therefore, in summary the window of modes for which the black hole is adiabatically evolving while still being in the sonic (linearly dispersive) regime is $\lambda=0.3-1$ mm and this window overlaps with the set of efficiently superradiating modes controlled by $\omega_\text{super}$ and mode conversion at the horizon which is controlled by $\kappa$.

{\textit{Experiments.}}--- The experimental layout is shown in Fig.~\ref{fig:setup}: a broad CW laser beam with gaussian profile and vacuum wavelength $\lambda=532$ nm that is launched through a diffractive phase mask on fused silica glass substrate, that is designed such that the first diffracted order carries the desired phase $\phi=m\theta-2\pi\sqrt{r/r_0}$ with $m$=2 and $r_0=0.5$ mm.
\begin{figure}[t!]
\begin{center}
\includegraphics[width=0.48\textwidth]{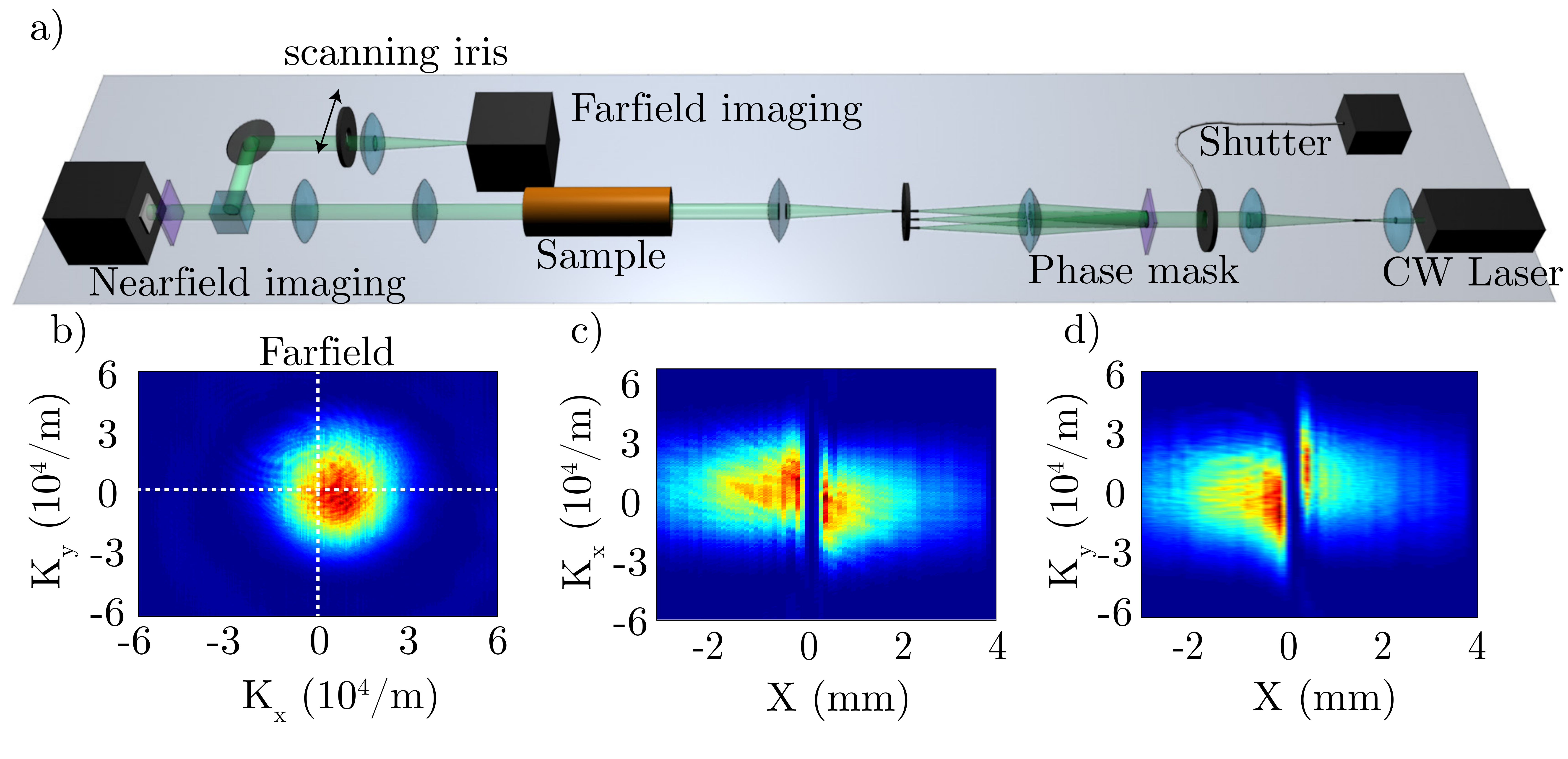}
\caption{a) Experimental setup: A CW 532 nm laser beam is launched onto a diffractive phase mask, that imprints the desired phase. Diffracted orders are selected by a spatial filter in the focus of a 4f-imaging telescope. A mechanical shutter shuts the beam on/off that is then launched through a 13 cm long tube filled with a nonlinear methanol graphene solution. The near-field and far-field intensity is imaged at the output facet of the sample onto a CCD camera. The spatially resolved far-field is recorded by selecting small areas of the beam profile by an iris ($\oslash\approx200$ $\mu$m) that is scanned across the beam diameter. b) Example of the spatially selected far-field. The location $(K_x(x,y=0),K_y(x,y=0))$ of the spot is tracked during the scan to obtain the flow velocity. c, d) Lineouts of a) along the $K_x$ and $K_y$ (dotted white crosshair) as a function of iris position $x$.}
\label{fig:setup}
\end{center}
\end{figure}
%
%
%
%
The beam is then imaged shortly after the phase mask by a 4f-imaging system onto the input facet of the nonlinear sample, so that the $+1$ and $-1$ orders can be selected by a pinhole in the farfield of the first lens, thus selecting an $m=2$, radial ingoing flow (black hole) or $m=-2$, outgoing flow (white hole). The sample consists of a cylindrical cell with length $L=13$ cm and radius $W=1$ cm filled with a dilute methanol/graphene solution as a nonlinear medium \cite{Vocke2015}. Finally, the near- and farfield intensity at the output facet of the nonlinear sample is recorded by a CCD camera. 
\begin{figure}[t!]
\begin{center}
\includegraphics[width=0.48\textwidth]{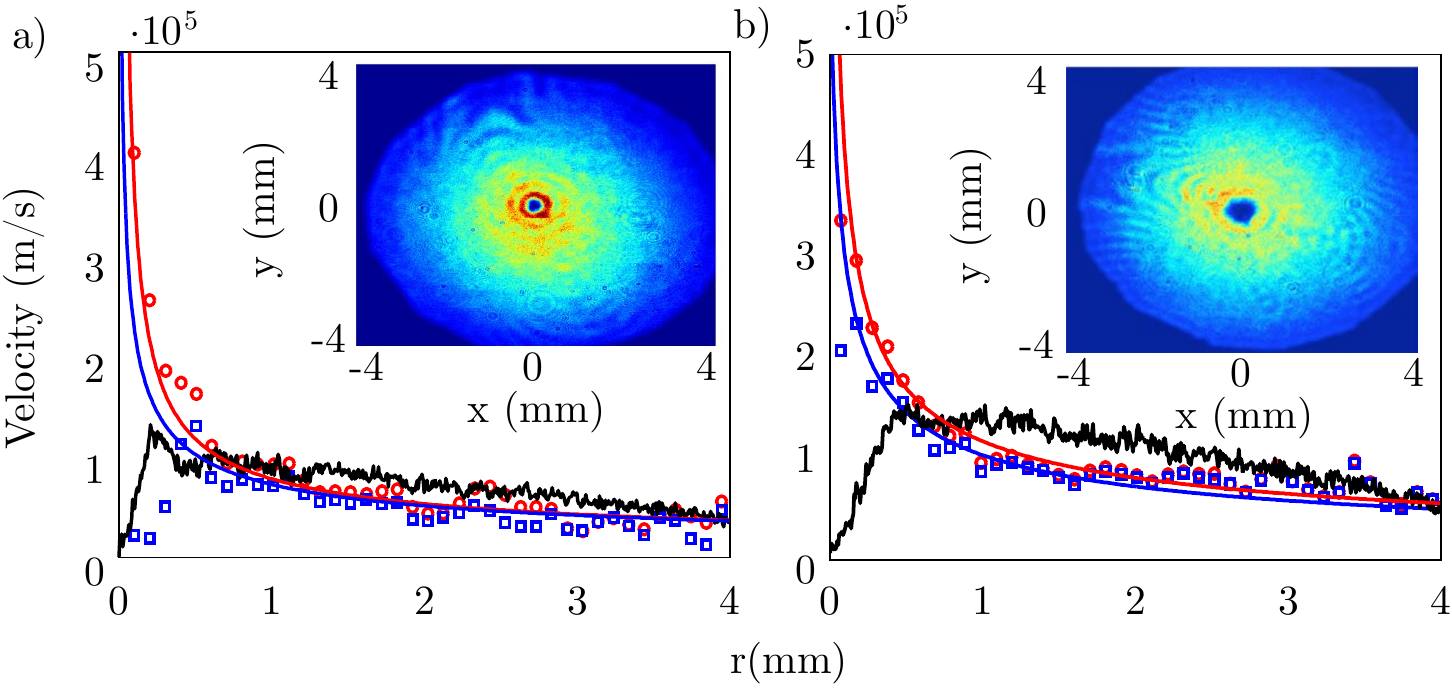}
\caption{a): Black hole with experimental parameters $r_0=0.5$ mm, $m=2$, $P=140$ mW, $t=200$ ms. b): White hole  with the same parameters and $t=600$ ms. Radial and total flow velocities (red circles and blue squares) and speed of sound (black).  An event horizon and ergosphere are found near $x\approx\pm0.5$ mm, where the $v_{\mathrm{tot}}(x)=c_s(x)$ and $v_{r}(x)=c_s(x)$. Inset: nearfield intensity distribution at the sample output. The color code shows the intensity from 0 (dark blue) to 1 $\mathrm{W/cm^2}$ (dark red). }
\label{fig:BHvelocities_exp}
\end{center}
\end{figure}
Methanol provides a thermal defocusing nonlinearity induced by absorbed laser power $\Delta n(r)=\beta\Delta T(r)$ with a thermo-optic coefficient $\beta=-4\times 10^{-4}$ 1/K while nanometric graphene flakes are added to increase absorption to $\approx 25\%$ along propagation in order to provide sufficient nonlinearity. As mentioned above, heat diffusion in the transverse directions, smears out the photon-photon interaction with a spatial extent described by the nonlocal length $\sigma$. The evolution time (taken for a given plane to propagate at the speed of light through the sample) in the photon fluid ($<1$ ns) is much faster than the thermal diffusion time ($\sim 1$ s) in the system: we therefore exploit the slow build up of the thermal nonlinearity to control the degree of nonlocality in our system. We have fully characterised the time dependent build up of the nonlinear parameters $\sigma(t)$ and $\gamma(t)$, thus precisely estimating the local sound speed $c_s(r,t)=\sqrt{c^2|\gamma|\rho(r,t)/n_0^3}$ as well as the nonlocal length  (see Appendix): by performing measurements after a short 200 ms time delay from the opening of the laser shutter and with an exposure time of 20 ms on the camera we effectively observe the evolution of the beam with a sufficiently high nonlinearity to ensure superfluid behaviour yet with a nonlocal length less than 166 $\mu$m, i.e. smaller than the healing length and thus can be ignored for the purpose of the experiments presented here.
\begin{figure}[t!]
\begin{centering}
\includegraphics[width=7cm]{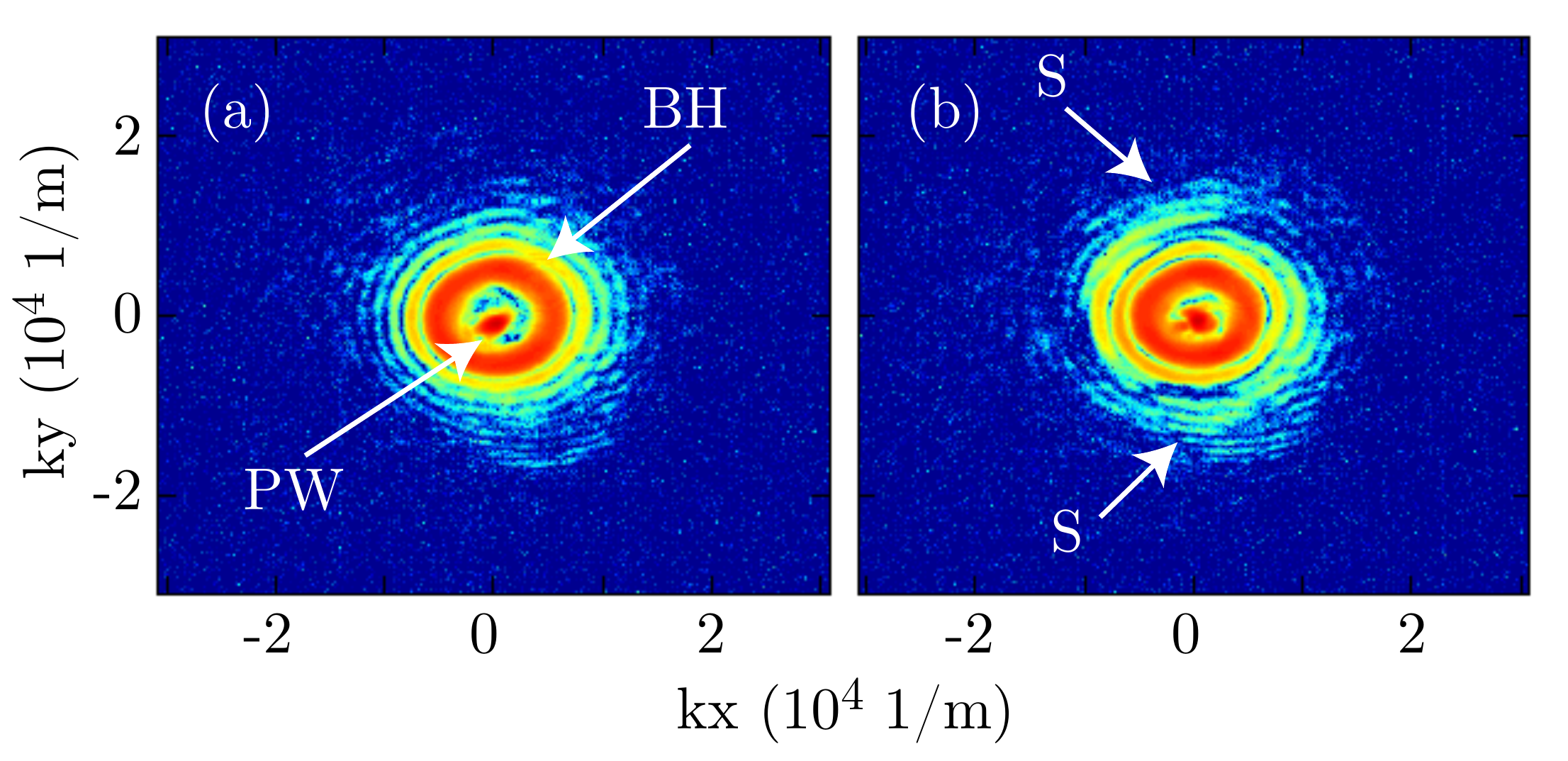}
\caption{(a) Far-field images of the black-hole, seen as a ring, indicated with `BH' together with a plane wave probe beam, indicated with `PW'. Image recorded at t=0: no interaction has occurred between the two. (b) At a later time ($t=300$ ms), scattering is observed of the PW from the BH, visible as spiralling tails, indicated with `S'.}
\label{scatter}
\end{centering}
\end{figure}
The photon fluid flow is proportional to the gradient of the spatial phase and can be expressed as:
\begin{equation}
	\begin{pmatrix} v_x(r)\\ v_y(r) \end{pmatrix}=\frac{c}{n_0k_0}\nabla\phi(r)=\frac{c}{n_0k_0}\begin{pmatrix} K_x(r)\\ K_y(r) \end{pmatrix} 
\end{equation}
This expression highlights that the local flow can be measured by recording the spatially resolved components $K_{x,y}(r)$ in the far-field. 
This is performed by isolating points of the beam in the near-field with an aperture ($\oslash\approx200$ $\mu$m) and measuring the location $(K_x(r),K_y(r))$ of the intensity peaks in the far-field (focal plane of a lens). 
The absolute values of the measured flows as well as the speed of sound are shown in Fig.~\ref{fig:BHvelocities_exp}a) for a rotating black hole and Fig.~\ref{fig:BHvelocities_exp}b) for a rotating white hole. An acoustic horizon can be identified in both cases at the intersection of the radial flow $v_r$ (blue squares) and speed of sound $c_s$  (solid black line) at $r\approx$0.5 mm. The ergosphere is found at a larger radius than the horizon where the total flow $v_{tot}^2=v_r^2+v_\theta^2$ intersects the sound speed. 
These measurements therefore completely identify the full spacetime metric of the rotating photon fluid.\\
{Finally,  we show a preliminary measurement with the sole purpose of supporting the idea that one may indeed observing scattering processes between a weak probe beam and a BH vortex background. Figure~\ref{scatter}(a) shows the far-field of the cell output for $t=0$, (i.e. immediately as the laser is switched on, thus not allowing time for the nonlinearity and interaction to build up) where the vortex is seen as ring (indicated with `BH') and the overlapping plane-wave probe (at a small angle) beam is seen as a spot (indicated with `PW'). At longer times ($t=$300 ms), the nonlinearity has built up, interaction ensues and scattering of the probe from the BH is seen in Fig.~\ref{scatter}(b) as a spiral in the outer region (indicated with `S).}

{\textit{Conclusions.}}---We have shown that by shaping the input phase profile of a photon fluid that it is possible to create a rotating black-hole-like spacetime metric that is to a good approximation stationary over relatively long propagation lengths. By fully characterising the spacetime metric, we were able to identify the location of the horizon and ergosphere, i.e. the regions where sound waves cannot escape the radial flow and where sound waves cannot counter-rotate with respect to the vortex flow, respectively. Superfluid behaviour and hence direct interpretation of wave evolution on the spacetime background in terms of the Klein-Gordon equation for a massless scalar field is seen to ensue over a broad range of wavelengths. This therefore holds promise for experiments aimed at investigating important scattering effects such as Penrose superradiance, here enriched by the unique feature that the spacetime vorticity is also quantised. Penrose superradiance is usually described as an energy extraction mechanism resulting in spin-down of the black hole: an intriguing question then is how this process ensues in the case of a quantised vortex flow.  

\section*{Appendix}

The refractive index change of the dilute methanol/graphene solution is directly proportional to the temperature change inside the material, i.e. $\Delta n(r)=\beta\Delta T(r)$, induced by heat of the absorbed laser power. If the absorption is low, heat flow along the propagation direction can be neglected and the thermal nonlinearity can be accurately modelled by the analytical solution of the steady state 2D heat equation using a distributed loss model \cite{Vocke2015a}. In previous works we have shown that at steady state the thermal nonlinearity in our system is well described by $\Delta n(r)=\gamma\int dr'R(r-r')|E(r')|^2$ with the nonlocal response function $R(r-r')=1/(2\pi\sigma^2)K_0(|r-r'|/\sigma)$ and $K_0$ the zeroth-order modified Bessel function of the second kind \cite{Vocke2015a,Roger2016}. Here we apply this model to the time dependent build up of the nonlinearity, i.e:
\begin{equation} 
\Delta n(r,t)=\gamma(t)\int dr'R(r-r',\sigma(t))|E(r')|^2
\label{eq:Deltan}
\end{equation}
that is used to describe the non-stationary change of refractive index after the laser is switched on. We verified that the spatial profiles at all times are to very good approximation described by the steady state solution with a time dependent nonlocal length $\sigma(t)$ and nonlinear coefficient $\gamma(t)$.
The characterisation was conducted by a method reported by Minovich et al. \cite{Minovich2007}, that allows to measure the nonlinear refractive index profile induced by a narrow gaussian laser beam. The induced nonlinear refractive index change in the transverse plane to propagation $\Delta n(x,y,t)$ can be visualised by a spatial phase change $\Delta\Phi(x,y,t)$ of a weak probe beam propagating through the illuminated sample. 
\begin{figure}[t!]
\begin{center}
\includegraphics[width=0.48\textwidth]{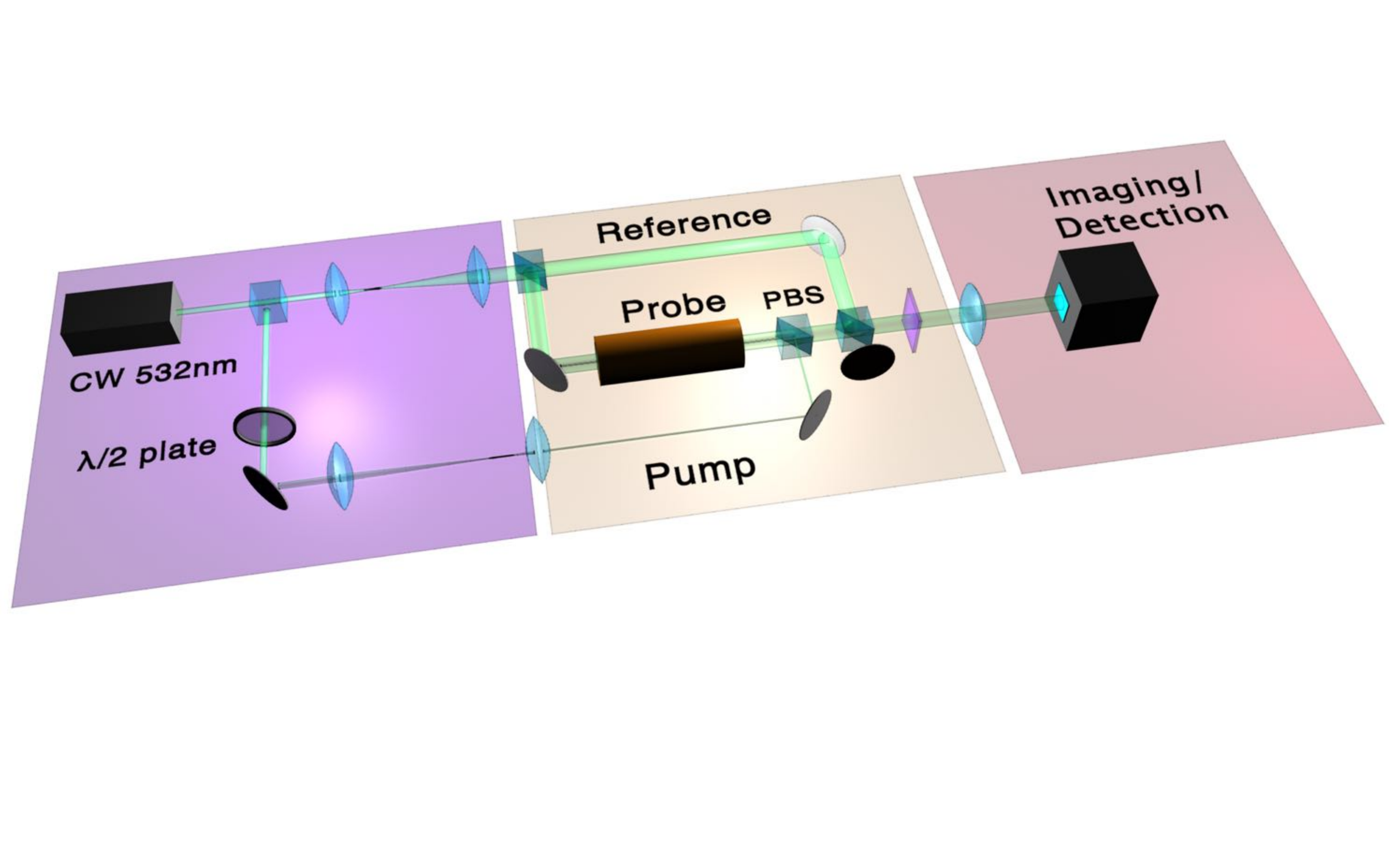}
\caption{Experimental setup: A monochromatic laser beam is split into 3 components: pump, probe and reference beam. The strong pump induces a transverse heat profile that can be measured by a phase shift in the probe beam profile with respect to the reference beam. The pump is cross-polarized and counter propagating through the nonlinear sample with respect to the other beams. The relative phase change $\Delta\Phi(x,y,t)$ is measured by phase shifting interferometry at the output facet that is imaged by a 4-f telescope onto the camera.}
\label{fig:setup}
\end{center}
\end{figure}
The relative phase $\Phi(x,y,t)$ between the probe and a reference beam is measured by a technique known as phase shifting interferometry, where a set of interferograms is recorded and processed in an algorithm described in \cite{Kinnstaetter1988} to obtain the phase. The index and the phase change are related through $\Delta n(x,y,t)=\frac{\Delta\Phi(x,y,t)\lambda}{2\pi L}$, where $\lambda$ is the laser wavelength and $L$ the length of the nonlinear sample.
The phase change $\Delta\Phi(x,y,t)$ is measured at the sample output with respect to an undistorted reference phase $\Phi_0$, i.e. taken at t=0 s. Examples of the spatial $\Delta n(x,y,t)$ profile for different laser powers are shown in Fig. \ref{fig:speedofsound_experiments}a). The temporal build up of the peak $\Delta n(x,y,t)$ is shown in Fig. \ref{fig:speedofsound_experiments}b)-c). In the time frame below one second, $\Delta n(x,y,t)$ is a linear function of the used laser power and intensity. The linearity is crucial, since it allows to precisely control the nonlinearity through scaling the laser power in the flow velocity experiment, where the intensities are considerably smaller ($\approx 0.7$ $\mathrm{W/cm}^2$) than the intensities used here ($\approx 30-90$ $\mathrm{W/cm}^2$). The spatial profiles $\Delta n(x,y,t)$ were fitted with Eq. (\ref{eq:Deltan}) using $\sigma(t)$ and $\gamma(t)$ as independent parameters and the results are shown in Fig. \ref{fig:speedofsound_experiments}d)-e). The nonlinear coefficient $|\gamma(t)|$ is well described by a power law $|\gamma(t)|=\gamma_0\times(t/t_0)^b$ in the time window below 10s with $\gamma_0=\gamma(t_0=1s)=1.45\times 10^{-6}$ $\mathrm{cm^2/W}$ and $b=0.74$. With this, we can precisely measure the local sound speed $c_s(r,t)=\sqrt{c^2|\gamma(t)|\rho(r)/n_0^3}$ by recording the laser intensity at the sample output.
\begin{figure*}[ht!]
\begin{center}
\includegraphics[width=0.9\textwidth]{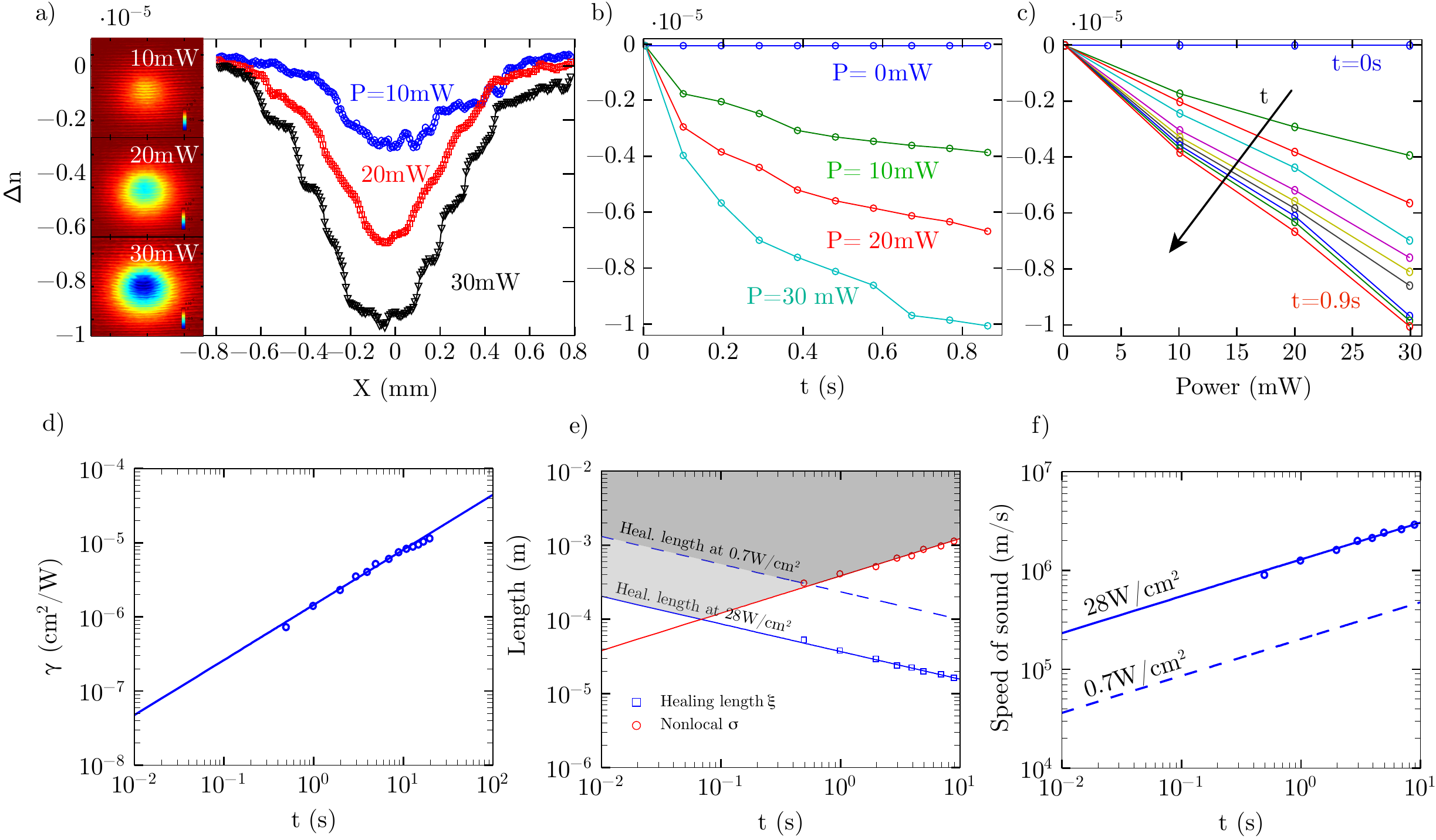}
\caption{Time-dependent thermal nonlinearity in methanol/graphene for short times ($<1$ s) and different pump powers with a pump beam waist radius $w=131$ $\mu$m. a) Lineouts $\Delta n(x,y=0,t=0.9 s)$ of thermally induced refractive index profiles after 0.9 s at P=10-30 mW. Insets: Corresponding 2D contour plots  b)-c) Peak nonlinear refractive index $\Delta n(x=0,y=0,t)$ as a function of time and power. $\Delta n$ is linearly proportional to the used power in the measured time window. d) Local nonlinear coefficient $|\gamma(t)|$ derived from fitting Eq. \ref{eq:Deltan} to measured $\Delta n(x,y,t)$. e) Time dependent photon fluid parameters. The dashed lines are calculated by linear scaling according to the intensity ratios. The grey shaded areas mark the wavelengths of excitations that travel with constant sound speed and have superfluid phonon character. f) time dependent speed of sound.}
\label{fig:speedofsound_experiments}
\end{center}
\end{figure*} The nonlocal length $\sigma(t)$ follows $\sigma(t)=\sigma_0\sqrt{t/t_0}$ with $\sigma_0=385$ $\mu$m which is in good agreement to $\sigma_0=\sqrt{\frac{\kappa}{\rho_0C}}=321\mu$m derived from the time-dependent 2D heat equation and using material parameters for methanol ($\rho_0=793$ kg/m$^3$, $C_p=2.5$ J/(gK), $\kappa=0.204$ W/mK). The nonlocal length is power independent and only depends on the boundary conditions \cite{Vocke2015a}. The flow speed measurements shown in the main manuscript were conducted at t=200ms, yielding a nonlinear coefficient of $|\gamma(t=0.2s)|=4.4\times 10^{-7}$ $\mathrm{cm}^2/W$ and $\sigma(t=0.2)=166$ $\mu$m.
The time dependent healing length and nonlocal length that define the hydrodynamic limit are plotted for two intensities in Fig. \ref{fig:speedofsound_experiments}e).  The grey shaded areas denote the regimes where the photon fluid is a superfluid with a constant speed of sound for excitations with wavelength larger than both parameters. Thus, between t=0.1s and 1s waves larger than $\Lambda\approx400$ $\mu$m propagate at a constant speed and therefore the definition of an acoustic horizon makes sense.

{\textit{Acknowledgements}}--- D.F. and A.P. acknowledge financial support from the European Research Council under the European Union Seventh Framework Programme (FP/2007-2013)/ERC GA 306559, Marie Sk\l odowska-Curie grant agreement No. 659301 and EPSRC (UK, Grant No. EP/M009122/1).

\newpage
%
\end{document}